# A Hierarchical Approach for Dependability Analysis of a Commercial Cache-Based RAID Storage Architecture

M. Kaâniche[1*], L. Romano[2†], Z. Kalbarczyk[2], R. Iyer[2] and R. Karcich[3]

[2]Center for Reliable and High-Performance Computing
University of Illinois at Urbana-Champaign
1308 W. Main St., Urbana, IL 61801, USA
lrom@grid.unina.it; {kalbar, yer}@crhc.uiuc.edu

[1]LAAS-CNRS,
7, Av. du Colonel Roche
31077 Toulouse Cedex 4
France
kaaniche@laas.fr

[3]Storage Technology
2270 S 88th St. MS 2220
Louisville, CO 80028,
USA
karciRM@louisville.stortek.com

**Abstract**

*We present a hierarchical simulation approach for the dependability analysis and evaluation of a highly available commercial cache-based RAID storage system. The architecture is complex and includes several layers of overlapping error detection and recovery mechanisms. Three abstraction levels have been developed to model the cache architecture, cache operations, and error detection and recovery mechanism. The impact of faults and errors occurring in the cache and in the disks is analyzed at each level of the hierarchy. A simulation submodel is associated with each abstraction level. The models have been developed using DEPEND, a simulation-based environment for system-level dependability analysis, which provides facilities to inject faults into a functional behavior model, to simulate error detection and recovery mechanisms, and to evaluate quantitative measures. Several fault models are defined for each submodel to simulate cache component failures, disk failures, transmission errors, and data errors in the cache memory and in the disks. Some of the parameters characterizing fault injection in a given submodel correspond to probabilities evaluated from the simulation of the lower-level submodel. Based on the proposed methodology, we evaluate and analyze 1) the system behavior under a real workload and high error rate (focusing on error bursts), 2) the coverage of the error detection mechanisms implemented in the system and the error latency distributions, and 3) the accumulation of errors in the cache and in the disks.*

## 1 Introduction

A RAID (Redundant Array of Inexpensive Disks) is a set of disks (and associated controller) that can automatically recover data when one or more disks fail [4, 13]. Storage architectures using a large cache and RAID disks are becoming a popular solution for providing high performance at low cost without compromising much data reliability [5, 10]. The analysis of these systems is focused on performance (see e.g., [9, 11]). The cache is assumed to be error free, and only the impact of errors in the disks is investigated. The impact of errors in the cache is addressed (to a limited extent) from a design point of view in [12], where the architecture of a fault-tolerant, cache-based RAID controller is presented. Papers studying the impact of errors in caches can be found in other applications not related to RAID systems (e.g., [3]).

In this paper, unlike previous work, which mainly explored the impact of caching on the performance of disk arrays, we focus on dependability analysis of a cache-based RAID controller. Errors in the cache might have a significant impact on the performance and dependability of the overall system. Therefore, in addition to the fault tolerance capabilities provided by the disk array, it is necessary to implement error detection and recovery mechanisms in the cache. This prevents error propagation from the cache to the disks and users, and it reduces error latency (i.e., time between the occurrence of an error and its detection or removal). The analysis of the error detection coverage of these mechanisms, and of error latency distributions, early in the design process provides valuable information. System manufacturers can understand, early on, the fault tolerance capabilities of the overall design and the impact of errors on performance and dependability.

In our case study, we employ hierarchical simulation, [6], to model and evaluate the dependability of a commercial cache-based RAID architecture. The system is decomposed into several abstraction levels, and the impact of faults occurring in the cache and the disk array is evaluated at each level of the hierarchy. To analyze the system under realistic operational conditions, we use real input traces to drive the simulation. The system model is based on the specification of the RAID architecture, i.e., we do not evaluate a prototype system. Simulation experiments are conducted using the DEPEND environment [7].

The cache architecture is complex and consists of sev-eral layers of overlapping error detection and recovery mechanisms. Our three main objectives are 1) to analyze how the system responds to various fault and error scenarios, 2) to analyze error latency distributions taking into account the origin of errors, and 3) to evaluate the coverage of error detection mechanisms. These analyses require a detailed evaluation of the system's behavior in the presence of faults. In general, two complementary approaches can be used to make these determinations: analytical modeling and simulation. Analytical modeling is not appropriate here, due to the complexity of the RAID architecture. Hierarchical simulation offers an efficient method to conduct a detailed analysis and evaluation of error latency and error detection coverage using real workloads and realistic fault scenarios. Moreover, the analysis can be completed within a reasonable simulation time.

---

[*] Was a Visiting Research Assistant Professor at CRHC, on leave from LAAS-CNRS, when this work was performed.

[†] Was a Visiting Research Scholar at CRHC, on leave from Dipartimento di Informatica e Sistemistica, University of Naples, Italy

To best reproduce the characteristics of the input load, a real trace file, collected in the field, is used to drive the simulation. The input trace exhibits the well-known track skew phenomenon, i.e., a few tracks among the addressable tracks account for most of the I/O requests. Since highly reliable commercial systems commonly tolerate isolated errors, our study focuses on the impact of multiple near-coincident errors occurring during a short period of time (error bursts), a phenomenon which has seldom been explored. We show that due to the high frequency of system operation, a transient fault in a single system component can result in a burst of errors that propagate to other components. In other words, what is seen at a given abstraction level as a single error becomes a burst of errors at a higher level of abstraction. Also, we analyze how bursts of errors affect the coverage of error detection mechanisms implemented in the cache and how they affect the error latency distributions, (taking into account where and when the errors are generated). In particular, we demonstrate that the overlapping of error detection and recovery mechanisms provides high error detection coverage for the overall system, despite the occurrence of long error bursts. Finally, analysis of the evolution of the number of faulty tracks in the cache memory and in the disks shows an increasing trend for the disks but an almost constant number for cache memory.

This paper contains five sections. Section 2 describes the system architecture and cache operations, focusing on error detection and recovery mechanisms. Section 3 outlines the hierarchical modeling approach and describes the hierarchical model developed for the system analyzed in this paper. Section 4 presents the results of the simulation experiments. Section 5 summarizes the main results of the study and concludes the paper.

## 2 System presentation

The storage architecture analyzed in this paper (Figure 1) is designed to support a large amount of disk storage and to provide high performance and high availability. The storage system supports a RAID architecture composed of a set of disk drives storing data, parity, and Reed-Solomon coding information, which are striped across the disks [4]. This architecture tolerates the failure of up to two disks. If a disk fails, the data from the failed disk is reconstructed on-the-fly using the valid disks; the reconstructed data is stored on a hot spare disk without interrupting the service. Data transfer between the hosts and the disks is supervised by the array controller. The array controller is composed of a set of control units. The control units process user requests received from the channels and direct these requests to the cache subsystem. Data received from the hosts is assembled into tracks in the cache. The number of tracks corresponding to a single request is application dependent. Data transfers between the channels and the disks are performed by the cache subsystem via reliable and high-speed control and data busses. The cache subsystem consists of 1) a cache controller organized into cache controller interfaces to the channels and the disks and cache controller interfaces to the cache memory (these interfaces are made of redundant components to ensure a high level of availability) and 2) cache volatile and nonvolatile memory. Communication between the cache controller interfaces and the cache memory is provided by redundant and multidirectional busses (denoted as Bus 1 and Bus 2 in Figure 1). The cache volatile memory is used as a data staging area for read and write operations. The battery-backed nonvolatile memory is used to protect critical data against failures (e.g., data modified in the cache and not yet modified in the disks, information on the file system that is necessary to map the data processed by the array controller to physical locations on the disks).

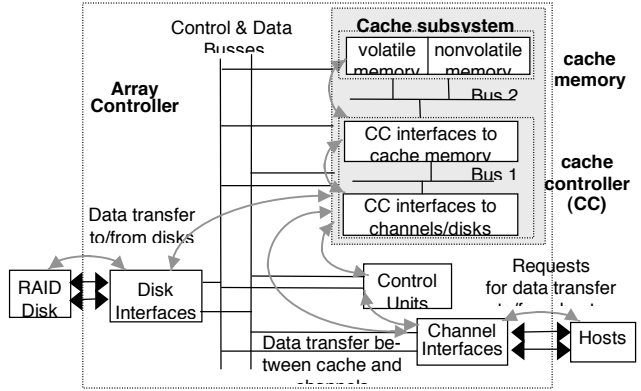

Figure 1: Array controller architecture, interfaces and data flow

### 2.1 Cache subsystem operations

The cache subsystem is for caching read and write requests. A track is always staged in the cache memory as a whole, even in the event of a write request involving only a few blocks of the track. In the following, we describe the main cache operations assuming that the unit of data transfer is an entire track.

*Read operation.* First, the cache controller checks for the requested track in the cache memory. If the track is already there («cache hit»), it is read from the cache and the data is sent back to the channels. If not («cache miss»), a request is issued to read the track from the disks and swap it to the cache memory. Then, the track is read from the cache.

*Write operation.* In the case of a cache hit, the track is modified in the cache and flagged as «dirty.» In the case of a cache miss, a memory is allocated to the track and the track is written into that memory location. Two write strategies can be distinguished: 1) write-through and 2) fast write. In the write-through strategy, the track is first written to the volatile memory. The write operation completion is signaled to the channels after the track is written to the disks. In the fast-write strategy, the track is written to the volatile memory and to nonvolatile memory. The write operation completion is signaled immediately. The modified track is later written to the disks according to a write-back strategy, which consists of transferring the dirty tracks to the disks, either periodically or when the amount of dirty tracks in the cache exceeds a predefined threshold. Finally, when space for a new track is needed in the cache, the track-replacement algorithm based on the Least-Recently-Used (LRU) strategy is applied to swap out a track from the cache memory.

*Track transfer inside the cache.* The transfer of a track between the cache memory, the cache controller, and the channel interfaces is composed of several elementary data transfers. The track is broken down into several data blocks to accommodate the parallelism of the different devices involved in the transfer. This also makes it possible to overlap several track transfer operations over the data busses inside the cache subsystem. Arbitration algorithms are implemented to synchronize these transfers and avoid bus hogging by a single transfer.

## 2.2 Error detection mechanisms

The cache is designed to detect errors in the data, address, and control paths by using, among other techniques, parity, error detection and correction codes (EDAC), and cyclic redundancy checking (CRC). These mechanisms are applied to detect errors in the data path in the following ways:

**Parity.** Data transfers, over Bus 1 (see Figure 1) are covered by parity. For each data symbol (i.e., data word) transferred on the bus, parity bits are appended and passed over separate wires. Parity is generated and checked in both directions. It is not stored in the cache memory but is stripped after being checked.

**EDAC.** Data transfers over Bus 2 and the data stored in the cache memory are protected by an error detection and correction code. This code is capable of correcting on-the-fly all single and double bit errors per data symbol and detecting all triple bit data errors.

**CRC.** Several kinds of CRC are implemented in the array controller. Only two of these are checked or generated

| Error Location | Error detection mechanism | | | |
|---|---|---|---|---|
| | FE-CRC | Parity | EDAC | PS-CRC |
| Transfer: channel to cache | x | | | |
| CCI to channels/disks | x | | | |
| Bus 1 | x | x | | |
| CCI to cache memory | x | | | |
| Bus 2 | x | | x | |
| Cache memory | x | | x | |
| Transfer: cache to disk | x | | | x |
| Disks | x | | | x |
| *Error detection condition* | < 4 ds with errors | odd # of errors per ds | < 4 bit-errors per ds | < 4 ds with errors |

ds= data symbol, CCI = Cache Controller Interface
Table 1. Error detection efficiency with respect to the location and the number of errors

within the cache subsystem: the frontend CRC (FE-CRC) and the physical sector CRC (PS-CRC). FE-CRC is appended, by the channel interfaces, to the data sent to the cache during a write request. It is checked by the cache controller. If FE-CRC is valid, it is stored with the data in the cache memory. Otherwise, the operation is interrupted and a CRC error is recorded. FE-CRC is checked again when a read request is received from the channels. Therefore, extra-detection is provided to recover from errors that may have occurred while the data was in the cache or in the disks, errors that escaped the error detection mechanisms implemented in the cache subsystem and the disk array. PS-CRC is appended by the cache controller to each data block to be stored in a disk sector. The PS-CRC is stored with the data until a read from disk operation occurs. At this time, it is checked and stripped before the data is stored in the cache. The same algorithm is implemented to compute FE-CRC and PS-CRC. This algorithm guarantees detection of three or fewer data symbols in error in a data record.

Table 1 summarizes the error detection conditions for each mechanism presented above, taking into account the component in which the errors occur and the number of noncorrected errors occurring between the computation of the code and its being checked. The (x) symbol means that errors affecting the corresponding component can be detected by the mechanism indicated in the column. It is noteworthy that the number of check bits and the size of the data symbol (ds) mentioned in the error detection condition are different for parity, EDAC, and CRC.

## 2.3 Error recovery and track reconstruction

Besides EDAC, which is able to automatically correct some errors by hardware, software recovery procedures are invoked when errors are detected by the cache subsystem. Recovery actions mainly consist of retries, memory fencing, and track-reconstruction operations. When errors are detected during a read operation from the cache volatile memory and the error persists after retries, an attempt is made to read the data from nonvolatile memory. If this op-eration fails, the data is read from the disk array. This operation succeeds if the data on the disks is still valid or it can be reconstructed (otherwise it fails). Figure 2 describes a simplified disk array composed of *n* data disks (D1 to Dn) and two redundancy disks (P and Q). Each row of the redundancy disks is computed based on the corresponding data tracks. For example, the first rows in disks P (P[1;n]) and Q (Q[1;n]) are obtained based on the data tracks T1 to Tn stored in the disks D1 to Dn. This architecture tolerates the loss of two tracks in each row; this condition will be referred to as the *track reconstruction condition*. The tracks that are lost due to disk failures or corrupted due to bit-errors can be reconstructed using the valid tracks in the row, provided that the track reconstruction condition is satisfied; otherwise data is lost. More information about disk reconstruction strategies can be found in [8].

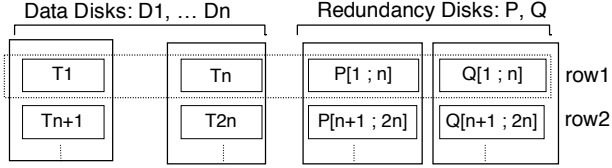

Figure 2: A simplified RAID

## 3 Hierarchical modeling methodology

We propose a hierarchical simulation approach to enable an efficient, detailed dependability analysis of the RAID storage system described in the previous section.

Establishing the proper number of hierarchical levels and their boundaries is not trivial. Several factors must be considered in determining an optimal hierarchical decomposition that provides a significant simulation speed-up with one minimal loss of accuracy: 1) system complexity, 2) the level of detail of the analysis and the dependability measures to be evaluated, and 3) the strength of system component interactions (weak interactions favor hierarchical decomposition).

In our study, we define three hierarchical levels (summarized in Figure 3) to model the cache-based storage system. At each level, the behavior of the shaded components is detailed in the lower-level model. Each model is built in a modular fashion and is characterized by:

- the components to be modeled and their behavior,
- a workload generator specifying the input distribution,
- a fault dictionary specifying the set of faults to be injected in the model, the distribution characterizing the occurrence of faults, and the consequences of the fault with the corresponding probability of occurrence, and
- the outputs derived from the submodel simulation.

For each level, the workload can be a real I/O access trace or generated from a synthetic distribution (in this study we use a real trace of user I/O requests). The effects of faults injected at a given level are characterized by statistical distributions (e.g., probability and number of errors occurring during data transfer inside the cache). Such distributions are used as inputs for fault injection at the next higher level. This mechanism allows the propagation of fault effects from lower-level models to higher-level models.

In the model described in Figure 3, the system behavior, the granularity of the data transfer unit, and the quantitative measures evaluated are refined from one level to another. In the Level 1 model, the unit of data transfer is a set of tracks to be read or written from a user file. In Level 2, it is a single track. In Level 3, the track is decomposed into a set of data blocks, each of which is composed of a set of data symbols. In the following subsections, we describe the three levels. In this study, we address Level 2 and Level 3 models, which describe the internal behavior of the cache and RAID subsystems in the presence of faults. Level 1 is included to illustrate the flexibility of our approach. Using the hierarchical methodology, additional models can be built on top of Level 2 and Level 3 models to study the behavior of other systems relying on the cache and RAID subsystems.

### 3.1 Level 1 model

Level 1 model translates user requests to read/write a specified file into requests to the storage system to read/write the corresponding set of tracks. It then propagates the replies from the storage system back to the users, taking into account the presence of faults in the cache and RAID subsystems. A file request (read, write) results in a sequence of track requests (read, fast-write, write-through). Concurrent requests involving the same file may arrive from different users. Consequently, a failure in a track operation can affect multiple file requests. In the Level 1 model, the *cache subsystem* and the *disk array* are modeled as a single entity—a black box. A fault dictionary specifying the results of track operations is defined to characterize the external behavior of the black box in the presence of faults. There are four possible results for a track operation (from the perspective of occurrence, detection, and correction of errors): 1) successful read/write track operation (i.e., absence of errors, or errors detected and corrected), 2) errors detected but not corrected, 3) errors not detected, and 4) service unavailable. Parameter values representing the probability of the occurrence of these events are provided by the simulation of the Level 2 model. Two types of outputs are derived from the simulation of the Level 1 model: 1) quantitative measures characterizing the probability of user requests failure and 2) the workload distribution of read or write track requests received by the cache subsystem. This workload is used to feed the Level 2 model.

### 3.2 Level 2 model

The Level 2 model describes the behavior in the presence of faults of the cache subsystem and the disk array. Cache operations and the data flow between the cache controller, the cache memory, and the disk array are described to identify scenarios leading to the outputs described in the Level 1 model and to evaluate their probability of occurrence. At Level 2, the data stored in the cache memory and the disks is explicitly modeled and structured into a set of tracks. Volatile memory and nonvolatile memory are modeled as separate entities. A track transfer operation is seen at a high level of abstraction. A track is seen as an atomic piece of data, traveling between different subparts of the system (from user to cache, from cache to user, from disk to cache, from cache to disk), while errors are injected to the track and to the different components of the system. Accordingly, when a track is to be transferred between two communication partners, for example, from the disk to the cache memory, none of the two needs to be aware of the disassembling, buffering, and reassembling procedures that occur during the transfer. This results in a significant simulation speedup, since the number of events needing to be processed is reduced dramatically.

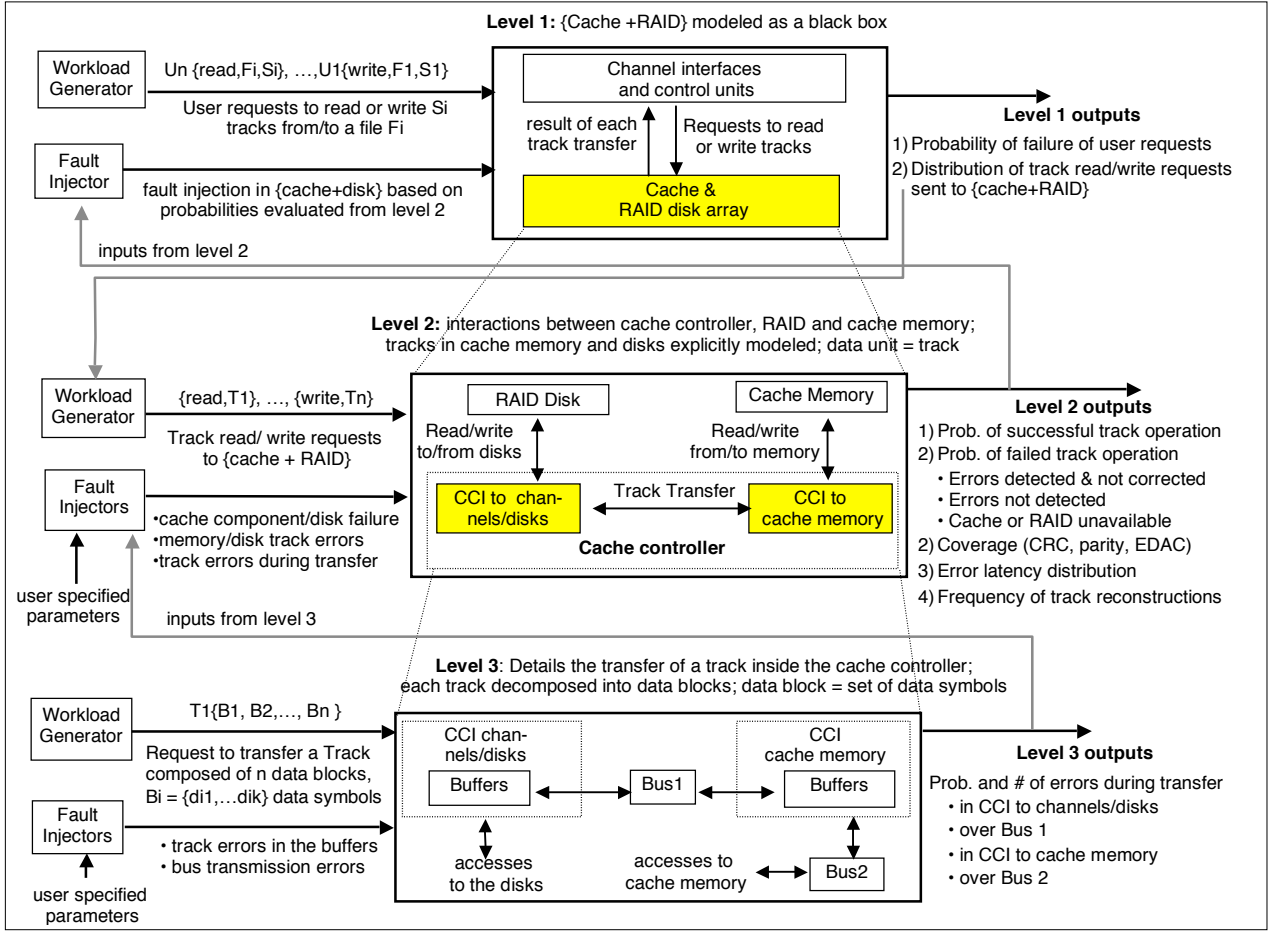

Figure 3: Hierarchical modeling of the cache-based storage system

**3.2.1 Workload distribution.** Level 2 model inputs correspond to requests to read or write tracks from the cache. Each request specifies the type of the access (read, write-through, fast-write) and the track to be accessed. The distribution specifying these requests and their interarrival times can be derived from the simulation of the Level 1 model, from real measurements (i.e., real trace), or by generating distributions characterizing various types of workloads.

**3.2.2 Fault models.** Specification of adequate fault models is essential to recreate realistic failure scenarios. To this end we distinguished three primary fault models, used to exercise and analyze error detection and recovery mechanisms of the target system. These fault models include 1) permanent faults leading to cache controller component failures, cache memory component failures, or disk failures, 2) transient faults leading to track errors affecting single or multiple bits of the tracks *while they are stored in the cache memory or in the disks*, and 3) transient faults leading to track errors affecting single or multiple bits *during the transfer* of the tracks by the cache controller to the cache memory or to the disks.

*Component failures.* When a permanent fault is injected into a cache controller component, the requests processed by this component are allocated to the other components of the cache controller that are still available. The reintegration of the failed component after repair does not interrupt the cache operations in progress. Permanent faults injected into a cache memory card or a single disk lead to the loss of all tracks stored in these components. When a read request involving tracks stored on a faulty component is received by the cache, an attempt is made to read these tracks from the nonvolatile memory or from the disks. If the tracks are still valid in the nonvolatile memory or in the disks, or if they can be reconstructed from the valid disks, then the read operation is successful, otherwise the data is lost. Note that when a disk fails, a hot spare is used to reconstruct the data and the failed disk is sent for repair.

*Track errors in the cache memory and the disks.* These correspond to the occurrence of single or multiple bit-errors in a track due to transient faults. Two fault injection strategies are distinguished: time dependent and load dependent. Time dependent strategy simulates faults occurring randomly. The time of injection is sampled from a predefined distribution, and the injected track, in the memory or in the disks, is chosen uniformly from the set of addressable tracks. Load dependent strategy aims at simulating the occurrence of faults due to stress. The fault injection rate depends on the number of accesses to the memory or to the disks (instead of the time), and errors are injected in the activated tracks. Using this strategy, frequently accessed tracks are injected more frequently than other tracks. For both strategies, errors are injected randomly into one or more bytes of a track. The fault injection rate is tuned to allow a single fault injection or multiple near-coincident fault injections (i.e., the fault rate is increased during a short period of time). This enables us to analyze the impact of isolated and bursty fault patterns.

*Track errors during transfer inside the cache.* Track errors can occur:
- in the cache controller interfaces with channels/disks before transmission over Bus 1 (see Figure 1), i.e., before parity or CRC computation or checking,
- during transfer over Bus 1, i.e., after parity computation,
- in the cache controller interfaces to cache memory before transmission over Bus 2, i.e., before EDAC computation, or
- during transfer over Bus 2, i.e., after EDAC computation.

To be able to evaluate the probability of occurrence and the number of errors affecting the track during the transfer, a detailed simulation of cache operations during this transfer is required. Including this detailed behavior in the Level 2 model would be far too costly in terms of computation time and memory occupation. For that reason, this simulation is performed in the Level 3 model. In the Level 2 model, a distribution is associated with each event described above, specifying the probability and the number of errors occurring during the track transfer. The track error probabilities are evaluated at Level 3.

**3.2.3 Modeling of error detection mechanisms.** Perfect coverage is assumed for cache components and disk failures due to permanent faults. The detection of track errors occuring when the data is in the cache memory or in the disks, or during the data transfer depends on (1) the number of errors affecting each data symbol to which the error detection code is appended and (2) when and where these errors occurred (see Table 1). The error detection modeling is done using a behavioral approach. The number of errors in each track is recorded and updated during the simulation. Each time a new error is injected into the track, the number of errors is incremented. When a request is sent to the cache controller to read a track, the number of errors affecting the track is checked and compared with the error detection conditions summarized in Table 1. During a write operation, the track errors that have been accumulated during the previous operations are overwritten, and the number of errors associated to the track is reset to zero.

**3.2.4 Quantitative measures.** Level 2 simulation enables us to reproduce several error scenarios and analyze the likelihood that errors will remain undetected by the cache or will cross the boundaries of several error detection and recovery mechanisms before being detected. Moreover, using the fault injection functions implemented in the model, we analyze (a) how the system responds to different error rates (especially burst errors) and input distributions and (b) how the accumulation of errors in the cache or in the disks and the error latency affect overall system behavior. Statistics are recorded to evaluate the following: coverage factors for each error detection mechanism, error latency distributions, and the frequency of track reconstruction operations. Other quantitative measures, such as the availability of the system and the mean time to data loss, can also be recorded.

**3.3 Level 3 model**

The Level 3 model details cache operations during the transfer of tracks from user to cache, from cache to user, from disk to cache, and from cache to disk. This allows us to evaluate the probabilities and number of errors occurring during data transfers (these probabilities are used to feed the Level 2 model, as discussed in Section 3.2). Unlike Level 2, which models a track transfer at a high level of abstraction as an atomic operation, in Level 3, each track is decomposed into a set of data blocks, which are in turn broken down into data symbols (each one corresponding to a predefined number of bytes). The transfer of a track is performed in several steps and spans several cycles. CRC, parity or EDAC bits are appended to the data transferred inside the cache or over the busses (Bus 1 and Bus 2). Errors during the transfer may affect the data bits as well as the check bits. At this level, we assume that the data stored in the cache memory and in the disk array is error free, as the impacts of these errors are considered in the Level 2 model. Therefore, we need to model only the cache controller interfaces to the channels/disks and to the cache memory and the data transfer busses. The Level 3 model input distribution defines the tracks to be accessed and the interarrival times between track requests. This distribution is derived from the Level 2 model.

Cache controller interfaces include a set of buffers in which the data to be transmitted to or received from the busses is temporarily stored (data is decomposed or assembled into data symbols and redundancy bits are appended or checked). In the Level 3 model, only transient faults are injected to the cache components (buffers and busses). During each operation, it is assumed that a healthy component will perform its task correctly, i.e., it will execute the operation without increasing the number of errors in the data it is currently handling. For example, the cache controller interfaces will successfully load their own buffers, unless they are affected by errors while performing the load operation. Similarly, Bus 1 and Bus 2 will transfer a data symbol and the associated information without errors, unless they are faulty while doing so. On the other hand, when a transient fault occurs, single or multiple bit-flips are continuously injected (during the transient) into the data symbols being processed. Since a single track transfer is a sequence of operations spanning several cycles, single errors due to transients in the cache components may lead to a burst of errors in the track currently being transferred. Due to the high operational speed of the components, even a short transient (a few microseconds) may result in an error burst, which affects a large number of bits.

## 4 Simulation experiments and results

In this section, we present the simulation results obtained from Level 2 and Level 3 to highlight the advantages of using a hierarchical approach for system dependability analysis. We focus on the behavior of the cache and the disks when the system is stressed with error bursts. Error bursts might occur during data transmission over busses, in the memory and the disks as observed, e.g., in [2]. It is well known that the CRC and EDAC error detection mechanisms provide high error detection coverage of single bit errors. Previously the impact of error bursts has not been extensively explored. In this section, we analyze the coverage of the error detection mechanisms, the distribution of error detection latency and error accumulation in the cache memory and the disks, and finally the evolution of the frequency of track reconstruction in the disks.

### 4.1 Experiment set-up

*Input distribution.* Real traces of user I/O requests were used to derive inputs for the simulation. Information provided by the traces included tracks processed by the cache subsystem, the type of the request (read, fast-write, write-through), and the interarrival times between the requests. Using a real trace gave us the opportunity to analyze the system under a real workload. The input trace described accesses to more than 127,000 tracks, out of 480,000 addressable tracks. As illustrated by Figure 4, the distribution of the number of accesses per track is not uniform. Rather a few tracks are generally more frequently accessed than the rest—the well-known track skew phenomenon. For instance, the first 100 most frequently accessed tracks account for 80% of the accesses in the trace; the leading track of the input trace is accessed 26,224 times, whereas only 200 accesses are counted for rank-100 track. The interarrival time between track accesses is about a few milliseconds, leading to high activity in the cache subsystem. Figure 5 plots the probability density function of the interarrival times between track requests. Regarding the type of the requests, the distribution is: 86% reads, 11.4% fast-writes and 2.6% write-through operations.

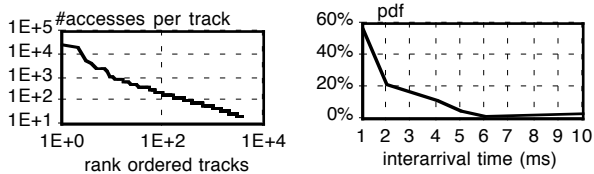

Figure 4: Track skew (Log-Log)   Figure 5: Interarrival time

*Simulation parameters.* We simulated a large disk array composed of 13 data disks and 2 redundancy disks. The RAID data capacity is 480,000 data tracks. The capacity of the simulated cache memory is 5% the capacity of the RAID. The rate of occurrence of permanent faults is $10^{-4}$ per hour for cache components (as is generally observed for hardware components) and $10^{-6}$ per hour for the disks [4]. The mean time for the repair of cache subsystem components is 72 hours (a value provided by the system manufacturer). Note that when a disk fails, a hot spare is used for the online reconstruction of the failed disk.

Transient faults leading to track errors occur more frequently than permanent faults. Our objective is to analyze how the system responds to high fault rates and bursts of errors. Consequently, high transient fault rates are assumed in the simulation experiment: 100 transients per hour over the busses, and 1 transient per hour in the cache controller interfaces, the cache memory and the disks. Errors occur more frequently over the busses than in the other components. Regarding the load-dependent fault injection strategy, the injection rate in the disk corresponds to one error each $10^{14}$ bits accessed, as observed in [4]. The same injection rate is assumed for the cache memory. Finally, the length of the error burst in the cache memory and in the disks is sampled from a normal distribution with a mean of 100 and a standard deviation of 10, whereas the length of the error burst during the track transfer inside the cache is evaluated from the Level 3 model as discussed in Section 3.3. The results presented in the following subsections correspond to the simulation of 24 hours of system operation.

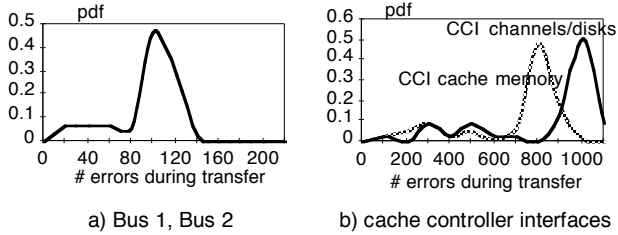

a) Bus 1, Bus 2    b) cache controller interfaces

Figure 6: Pdf of number of errors during track transfer given that a transient fault is injected

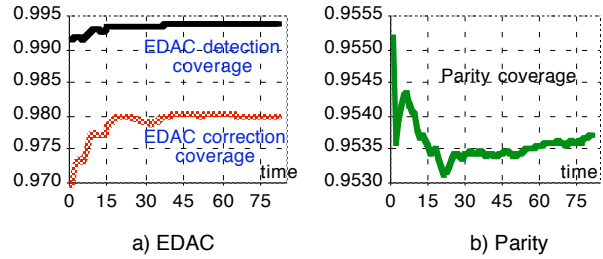

a) EDAC    b) Parity

Figure 7: EDAC and parity coverage during simulation time

### 4.2 Level 3 model simulation results

As discussed in Section 3.3, the Level 3 model aims at evaluating the number of errors occurring during the transfer of the tracks inside the cache due to transient faults over the busses and in the cache controller interfaces. We assumed that the duration of a transient fault is 5 microseconds. During the duration of the transient, single or multiple bit flips are continuously injected in the track data symbols processed during that time. The cache operational cycle for the transfer of a single data symbol is of the order of magnitude of a few nanoseconds. Therefore the occurrence of a transient fault might affect a large number of bits in a track. This is illustrated by Figure 6, which plots the conditional probability density function of the number of errors (i.e., number of bit-flips) occurring during the transfer over Bus 1 and Bus 2 (Figure 6-a) and inside the cache controller interfaces (Figure 6-b), given that a transient fault occurred. The distribution is the same for Bus 1 and Bus 2 due to the fact these busses have the same speed. The mean length of the error burst measured from the simulation is around 100 bits during transfer over the busses, 800 bits when the track is temporarily stored in the cache controller interfaces to cache memory, and 1000 bits when the track is temporarily stored in the cache controller interfaces to channels/disks. The difference between the results is related to the difference between the track transfer time over the busses and the track loading time inside the cache controller interfaces.

### 4.3 Level 2 model simulation results

We used the burst error distributions obtained from the Level 3 model simulation to feed Level 2 model as explained in Section 3.2. In following subsections we present and discuss the results obtained from the simulation of Level 2, specifically: 1) the coverage of the cache error detection mechanisms, 2) the error latency distribution, and 3) the error accumulation in the cache memory and disks and the evolution of the frequency of track reconstruction.

**4.3.1 Error detection coverage.** For all simulation experiments that we performed, the coverage factor measured for the frontend CRC and the physical sector CRC was 100%. This is due to the very high probability of detecting error patterns by the CRC algorithm implemented in the system (see Section 2.2). Regarding EDAC and parity, the coverage factors tend to stabilize as the simulation time increases (see Figures 7-a and 7-b, respectively). Each unit of time in Figures 7-a and 7-b corresponds to 15 minutes of system operation. Note that EDAC coverage remains high even though the system is stressed with long bursts occurring at a high rate, and more than 98% of the errors detected by EDAC are automatically corrected on-the-fly. This is due to the fact that errors are injected randomly in the track and the probability of having more than three errors in a single data symbol is low. (The size of a data symbol is around $10^{-3}$ the size of the track.) All the errors that escaped EDAC or parity have been detected by the frontend CRC upon a read request from the hosts. This result illustrates the advantages of storing the CRC with the data in the cache memory to provide extra detection of errors escaping EDAC and parity and to compensate for the relatively low parity coverage.

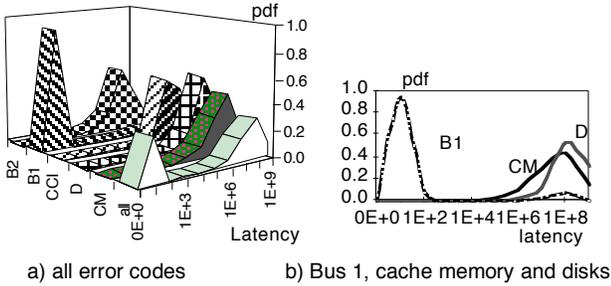

a) all error codes      b) Bus 1, cache memory and disks

Figure 8: Error latency distribution

#### 4.3.2 Error latency and error propagation.
When an error is injected in a track, the time of occurrence of the error and a code identifying which component caused the error are recorded. This allows us to monitor the error propagation in the system. Six error codes are defined: *CCI* (error occurred when data is stored in the cache controller interfaces to the channels/disks and to the cache memory), *CM* (error in the cache memory), *D* (error in the disk), *B1* (error during transmission over Bus 1), and *B2* (error during transmission over Bus 2). The time for an error to be overwritten (during a write operation) or detected (upon a read operation) is called *error latency*. Since a track is considered faulty as soon as an error is injected, we record the latency associated with the first error injected in the track. This means that the error latency that we measure corresponds to the time between when the track becomes faulty and when the errors are overwritten or detected. Therefore, the error latency measured for each track is the maximum latency for errors present in the track. Figure 8 plots the error latency probability density function for errors, as categorized above, and error latency for all samples without taking into account the origin of errors (the unit of time is 0.1 ms). The latter distribution is bimodal. The first mode corresponds to a very short latency that results mainly from errors occurring over Bus 1 and detected by parity. The second mode corresponds to longer latencies due to errors occurring in the cache memory or the disks, or to the propagation of errors occurring during data transfer inside the cache. Note that most of the errors escaping parity (error code B1) remain latent for a longer period of time (as discussed in Section 3.2.3).

The value of the latency depends on the input distribution. If the track is not frequently accessed, then errors present in the track might remain latent for a long period of time. Figure 8-b shows that the latency of errors injected in the cache memory is slightly lower than the latency of errors injected in the disk. This is because the disks are less frequently accessed than the cache memory. Finally, it is important to notice that the difference between the error latency distribution for error codes B1 and B2 (Figure 8-a) is due to the fact that data transfers over Bus1 (during read and write operations) are covered by parity, whereas errors occurring during write operations over Bus 2 are detected later by EDAC or by FE-CRC when the data is read from the cache. Consequently, it would be useful to check EDAC before data is written to the cache memory in order to reduce the latency of errors due to Bus 2.

#### 4.3.3 Error distribution in the cache memory and in the disks.
Analysis of error accumulation in the cache memory and disks provides valuable feedback, especially for scrubbing policy. Figure 9 plots the evolution in time of the percentage of faulty tracks in the cache memory and in disks (the unit of time is 15 minutes). An increasing trend is observed for the disks, whereas in the cache memory we observe a periodic behavior. In the latter case, the percentage of faulty tracks first increases and then decreases when either errors are detected upon read operations or are overwritten when tracks become dirty. Since the cache memory is accessed very frequently (every 5 milliseconds in average) and the cache hit rate is high (more than 60%), errors are more frequently detected and overwritten in the cache memory than in the disks. The increase of the number of faulty tracks in the cache affects the track reconstruction rate (number of reconstructions per unit of time), as illustrated in Figure 10. The average track reconstruction rate is approximately $8.7 \cdot 10^{-5}$ per millisecond. It is noteworthy that the detection of errors in the cache memory does not necessarily lead to the reconstruction of a track (the track might still be valid in the disks). Nevertheless, the detection of errors in the cache has an impact on performance due to the increase in the number of accesses to the disk. Figure 9 indicates that different strategies should be considered for disk and cache memory scrubbing. The disk should be scrubbed more frequently than the cache memory; this prevents error accumulation, which can lead to inability to reconstruct a faulty track.

## 5 Summary, discussion, and conclusions

The dependability of a complex and sophisticated cache-based storage architecture is modeled and simulated. To ensure reliable operation and to prevent data loss, the system employs a number of error detection mechanisms and recovery strategies, including parity, EDAC, CRC checking, and support of redundant disks for data reconstruction. Due to the complex interactions among these mechanisms, it is not a trivial task to accurately capture the behavior of the overall system in the presence of faults. To enable an efficient and detailed dependability analysis, we proposed a hierarchical, behavioral simulation-based approach in which the system is decomposed into several abstraction levels and a corresponding simulation model is associated with each level. In this approach, the impact of low-level faults is used in a higher level analysis. Using an appropriate hierarchical system decomposition, the complexity of individual models can be significantly reduced while preserving the model's ability to capture detailed system behavior. Moreover, additional details can be incorporated by introducing new abstraction levels and associated simulation models.

To demonstrate the capabilities of the methodology, we have conducted an extensive analysis of the design of a real, commercial cache RAID storage system. To our knowledge, this kind of analysis of a cache-based RAID system has not been accomplished either in academia or in the industry. The dependability measures used to characterize the system include coverage of the different error detection mechanisms employed in the system, error latency distribution classified according to the origin of an error, error accumulation in the cache memory and disks, and frequency of data reconstruction in the cache memory. To analyze the system under realistic operational conditions, we used real input traces to drive the simulations. It is important to emphasize that an analytical modeling of the system is not appropriate in this context due to the complexity of the architecture, the overlapping of error detection and recovery mechanisms, and the necessity of capturing the latent errors in the cache and the disks. Hierarchical simulation offers an efficient method to accomplish the above task and allows detailed analysis of the system to be performed using real input traces.

The specific results of the study are presented in the previous sections. It is, however, important to summarize the key points that demonstrate the usefulness of the proposed methodology. First, we focused on the analysis of the system behavior when it is stressed with high fault rates. In particular, we demonstrated that transient faults during a few microseconds—during data transfer over the busses or while the data is in the cache controller interfaces—may lead to bursts of errors affecting a large number of bits of the track. Moreover, despite the high fault injection rate, high EDAC and CRC error detection coverage was observed, and the relatively low parity coverage was compensated for by the extra detection provided by CRC, which is stored with the data in the cache memory.

The hierarchical simulation approach allowed us to perform a detailed analysis of error latency with respect to the origin of an error. The error latency distribution measured from the simulation, regardless the origin of the errors, is bimodal[†]: short latencies are mainly related to errors occurring and detected during data transfer over the bus protected by parity, and the highest error latency was observed for errors injected into the disks. The analysis of the evolution during the simulation of the percentage of faulty tracks in the cache memory and the disks showed that, in spite of a high rate of injected faults, there is no error accumulation in the cache memory, i.e., the percentage of faulty tracks in the cache varies within a small range (0.5% to 2.5%, see Section 4.3.3), whereas an increasing trend was observed for the disks (see Figure 9). This is related to the fact that the cache memory is accessed very frequently, and errors are more frequently detected and overwritten in the cache memory than in the disks. The primary implication of this result, together with the results of the error latency analysis, is the need for a carefully designed scrubbing policy capable of reducing the error latency with acceptable performance overhead. Simulation results suggest that the disks should be scrubbed more frequently than the cache memory in order to prevent error accumulation, which may lead to an inability to reconstruct faulty tracks.

We should emphasize that the results presented in this paper are derived from the simulation of the system using a single, real trace to generate the input patterns for the simulation. Additional experiments with different input traces and longer simulation times should be performed to confirm these results. Moreover, the results presented in this paper are preliminary, as we addressed only the impact of errors affecting the data. Continuation of this work will include modeling of errors affecting the control flow in cache operations. The proposed approach is flexible enough to incorporate these aspects of system behavior. Including control flow will obviously increase the complexity of the model, as more details about system behavior must be described in order to simulate realistic error scenarios and provide useful feed back for the designers. It is clear that this kind of detailed analysis cannot be done without the support of a hierarchical modeling approach.

## Acknowledgments

The authors are grateful to the anonymous reviewers whose comments helped improve the presentation of the paper and to Fran Baker for her insightful editing if our manuscript. This work was supported by the National Aeronautics and Space Administration (NASA) under grant NAG-1-613, in cooperation with the Illinois Computer Laboratory for Aerospace Systems and Software (ICLASS), and by the Advanced Research Projects Agency under grant DABT63-94-C-0045. The findings, opinions, and recommendations expressed herein are those of the authors and do not necessarily reflect the position or policy of the United States Government or the University of Illinois, and no official endorsement should be inferred.

## References

[1] J. Arlat, M. Aguera, Y. Crouzet, et al., «Experimental Evaluation of the Fault Tolerance of an Atomic Multicast System,» *IEEE Transactions on Reliability*, vol. 39, pp. 455-467, 1990.

[2] A. Campbell, P. McDonald, and K. Ray, «Single Event Upset Rates in Space,» *IEEE Transactions on Nuclear Science*, vol. 39, pp. 1828-1835, 1992.

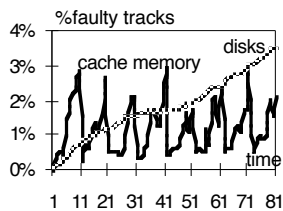

Figure 9: Percentage faulty tracks in cache and disks

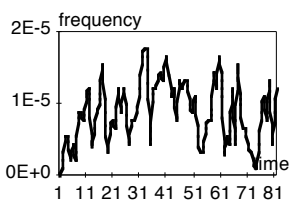

Figure 10: Frequency of track reconstruction

[†] Similar behavior was observed in other studies, e.g.,[1, 14].

[3] C.-H. Chen and A. K. Somani, «A Cache Protocol for Error Detection and Recovery in Fault-Tolerant Computing Systems,» *24th IEEE International Symposium on Fault-Tolerant Computing (FTCS-24),* Austin, Texas, USA, 1994, pp. 278-287.

[4] P. M. Chen, E. K. Lee, G. A. Gibson, et al., «RAID: High-Performance, Reliable Secondary Storage,» *ACM Computing Surveys*, vol. 26, pp. 145-185, 1994.

[5] M. B. Friedman, «The Performance and Tuning of a StorageTek RAID 6 Disk Subsystem,» *CMG Transactions*, vol. 87, pp. 77-88, 1995.

[6] K. K. Goswami, «Design for Dependability: A simulation-Based Approach,» PhD., University of Illinois at Urbana-Champaign, UILU-ENG-94-2204, CRHC-94-03, February 1994.

[7] K. K. Goswami, R. K. Iyer, and L. Young, «DEPEND: A simulation Based Environment for System level Dependability Analysis,» *IEEE Transactions on Computers*, vol. 46, pp. 60-74, 1997.

[8] M. Holland, G. Gibson, A., and D. P. Siewiorek, «Fast, On-Line Failure Recovery in Redundant Disk Arrays,» *23rd International Symposium on Fault-Tolerant Computing (FTCS-23),* Toulouse, France, 1993, pp. 422-431.

[9] R. Y. Hou and Y. N. Patt, «Using Non-Volatile Storage to improve the Reliability of RAID5 Disk Arrays,» *27th Int. Symposium on Fault-Tolerant Computing (FTCS-27), WA,* Seattle, 1997, pp. 206-215.

[10] G. E. Houtekamer, «RAID System: The Berkeley and MVS Perspectives,» *21st Int. Conf. for the Resource Management & Performance Evaluation of Enterprise Computing Systems (CMG'95),* Nashville, Tennesse, USA, 1995, pp. 46-61.

[11] J. Menon, «Performance of RAID5 Disk Arrays with Read and Write Caching,» *International Journal on Distributed and Parallel Databases*, vol. 2, pp. 261-293, 1994.

[12] J. Menon and J. Cortney, «The Architecture of a Fault-Tolerant Cached RAID Controller,» *20th Annual International Symposium on Computer Architecture,* San Diego, CA, USA, 1993, pp. 76-86.

[13] D. A. Patterson, G. A. Gibson, and R. H. Katz, «A Case for Redundant Arrays of Inexpensive Disks (RAID),» *ACM International Conference on Management of Data (SIGMOD),* New York, 1988, pp. 109-116.

[14] J. G. Silva, J. Carreira, H. Madeira, et al., «Experimental Assessment of Parallel Systems,» *26th International Symposium on Fault-Tolerant Computing (FTCS-26),* Sendai, Japan, 1996, pp. 415-424.